\documentclass[11pt]{article}
\usepackage{geometry}               
\usepackage{graphicx}
\usepackage{amssymb}
\usepackage{epstopdf}
\usepackage{url}
\newcommand{\be}{\begin{equation}}
\newcommand{\bea}{\begin{eqnarray}}
\newcommand{\beq}{\begin{equation}}

\newcommand{\ee}{\end{equation}}
\newcommand{\eea}{\end{eqnarray}}
\newcommand{\eeq}{\end{equation}}
\newcommand{\lsim}{\!\mathrel{\hbox{\rlap{\lower.55ex \hbox{$\sim$}} \kern-.34em \raise.4ex \hbox{$<$}}}}
\newcommand{\gsim}{\!\mathrel{\hbox{\rlap{\lower.55ex \hbox{$\sim$}} \kern-.34em \raise.4ex \hbox{$>$}}}}

\newcommand{\ltap}{\raisebox{-.4ex}{\rlap{$\sim$}} \raisebox{.4ex}{$<$}}   
\newcommand{\gtap}{\raisebox{-.4ex}{\rlap{$\sim$}} \raisebox{.4ex}{$>$}}   

\begin{document}
\begin{titlepage}
\flushright{RUNHETC-04-2007\\} \vspace{1in}
\begin{center}
{\Large \bf Broadening the Higgs Boson with   } \vspace{.07in}
{\Large \bf Right-Handed Neutrinos and a Higher Dimension Operator
at the Electroweak Scale}

\vspace{0.5in} {\bf Michael L. Graesser$^{1}$}

\vspace{.5cm}

{\it $^{1}$Department of Physics and NHETC,
 Rutgers University,
Piscataway, NJ 08540}

\vspace{0.2cm}

\end{center}
\vspace{0.8cm}
\begin{abstract}
The existence of certain TeV suppressed higher-dimension operators
may open up new decay channels for the Higgs boson to decay into
lighter right-handed neutrinos. These channels may dominate over all
other channels if the Higgs boson is light. For a Higgs boson mass
larger than $2 m_W$ the new decays are subdominant yet still of
interest. The right-handed neutrinos have macroscopic decay lengths
and decay mostly into final states containing leptons and quarks. A
distinguishing collider signature of this scenario is a pair of
displaced vertices violating lepton number. A general operator
analysis is performed using the minimal flavor violation hypothesis
to illustrate that these novel decay processes can occur while
remaining consistent with experimental constraints on lepton number
violating processes. In this context the question of whether these
new decay modes dominate is found to depend crucially on the
approximate flavor symmetries of the right-handed neutrinos.
\end{abstract}
\end{titlepage}

\section{Motivation}

Neutrino interactions with other Standard Model particles are
well-described, forming a cornerstone of the Standard Model itself.
But the origin of their masses remain unknown. If their masses are
generated by a local quantum field theory, then other degrees of
freedom must exist. These particles or ``right-handed neutrinos''
are either the missing Dirac partners of the neutrinos, or are much
heavier than the $O(\hbox{eV})$ mass scale of the neutrinos and
create a ``see-saw" mechanism. Which of these scenarios is realized
in Nature is dependent on the unknown scale of the Majorana mass
parameters of the right-handed neutrinos.

The existence of right-handed neutrinos may have other physical
consequences, depending on the size of their Majorana masses.
Right-handed neutrinos with Majorana masses violate overall lepton
number, which may have consequences for the origin of the observed
baryon asymmetry. Leptogenesis can occur from the out of equilibrium
decay of a right-handed neutrino with mass larger than the TeV scale
\cite{leptogenesis}. Interestingly, right-handed neutrinos with
masses below the electroweak scale may also lead to baryogenesis
\cite{nuSM}.

But if right-handed neutrinos exist, where did their mass come from?
The Majorana mass parameters are not protected by the gauge
invariance of the Standard Model, so an understanding of the origin
of their mass scale requires additional physics. The see-saw
mechanism with order unity Yukawa couplings prefers a large scale,
of order $10^{13-14}$ GeV. But in this case a new, intermediate
scale must be postulated in addition to the four mass scales already
observed in Nature. On the other hand, such a large scale might
occur naturally within the context of a Grand Unified Theory.

Here I explore the consequences of assuming that the Majorana
neutrino mass scale is generated at the electroweak scale
\footnote{For previous work on the phenomenology of electroweak
scale right-handed neutrinos, see
\cite{RHNEW1,RHNEW2,RHNEW3,RHNEW4}. None of these authors consider
the effects of TeV-scale suppressed higher dimension operators.}. To
then obtain the correct mass scale for the left-handed neutrinos
from the ``see-saw" mechanism, the neutrino Yukawa couplings must be
tiny, but not unreasonably small, since they would be comparable to
the electron Yukawa coupling. It might be natural for Majorana
masses much lighter than the Planck or Grand Unified scales to occur
in specific Randall-Sundrum type models \cite{rs} or their CFT dual
descriptions by the AdS/CFT correpsondance \cite{adscft}. But as the
intent of this paper is to be as model-independent as possible, I
will instead assume that it is possible to engineer electroweak
scale Majorana masses and use effective field theory to describe the
low-energy theory of the Higgs boson and the right-handed and
left-handed (electroweak) neutrinos. I will return to question of
model-building in the concluding section and provide a few
additional comments.

With the assumption of a common dynamics generating both the Higgs
and right-handed neutrino mass scales, one may then expect strong
interactions between these particles, in the form of higher
dimension operators. However since generic flavor-violating higher
dimension operators involving Standard Model fields and suppressed
only by the TeV are excluded, I will use throughout the minimal
flavor violation hypothesis
\cite{mfvprinciple,mfvprinciple2,wise:leptonviolation} in order to
suppress these operators. The purpose of this paper is to show that
the existence of operators involving the Higgs boson and the
right-handed neutrinos can significantly modify the phenomenology of
the Higgs boson by opening a new channel for it to decay into
right-handed neutrinos. I show that the right-handed neutrinos are
long-lived and generically have macroscopic decay lengths. For
reasonable values of parameters their decay lengths are anywhere
from fractions of a millimeter to tens of metres or longer if one of
the left-handed neutrinos is extremely light or massless. As they
decay predominantly into a quark pair and a charged lepton, a
signature for this scenario at a collider would be the observation
of two highly displaced vertices, each producing particles of this
type. Further, by studying these decays all the $CP$-preserving
parameters of the right-handed and left-handed neutrinos
interactions could be measured, at least in principle.

A number of scenarios for new physics at the electroweak scale
predict long-lived particles with striking collider features.
Displaced vertices due to long-lived neutral particles or kinks
appearing in charged tracks are predicted to occur in models of low
energy gauge mediation \cite{scott}. More recently models with a
hidden sector super-Yang Mills coupled weakly through a $Z'$ or by
mass mixing with the Higgs boson can produce dramatic signatures
with displaced jets or leptons and events with high multiplicity
\cite{strassler}. A distinguishing feature of the Higgs boson decay
described here is the presence of two displaced vertices where the
particles produced at each secondary vertex violate overall lepton
number.

That new light states or operators at the electroweak scale can
drastically modify Higgs boson physics has also been recently
emphasized. Larger neutrino couplings occur in a model with nearly
degenerate right-handed neutrino masses and vanishing tree-level
active neutrino masses, that are then generated radiatively at
one-loop \cite{RHNEW1}. Decays of the Higgs boson into a
right-handed and left-handed neutrino may then dominate over decays
to bottom quarks if the right-handed neutrinos are heavy enough.
Models of supersymmetry having pseudoscalars lighter than the
neutral Higgs scalar may have exotic decay processes for the Higgs
boson that can significantly affect limits and searches
\cite{cascadehiggs}. Supersymmetry without $R$-parity can have
striking new signatures of the Higgs boson \cite{dekaplan}. Two
common features between that reference and the work presented here
is that the Higgs boson decays into a 6-body final state and may be
discovered through displaced vertices, although the signatures
differ.

Interesting phenomena can also occur without supersymmetry. Adding
to the Standard Model higher dimension operators involving only
Standard Model fields can modify the Higgs boson production
cross-section and branching fractions \cite{manoharwise}. Such an
effect can occur in models with additional colored scalars coupled
to top quarks \cite{manoharwise2}.

The outline of the paper is the following. Section
\ref{sec:higgsdecay} discusses the new decay of the Higgs boson into
right-handed neutrinos. Section \ref{sec:naturalness} then discusses
various naturalness issues that arise in connection with the
relevant higher dimension operator. Section
\ref{sec:minimalflavorviolation} discusses predictions for the
coefficients of the new operator within the framework of minimal
flavor violation
\cite{mfvprinciple,mfvprinciple2,wise:leptonviolation}. It is found
that the predicted size of the higher dimension operators depends
crucially on the approximate flavor symmetries of the right-handed
neutrinos. How this affects the branching ratio for the Higgs boson
to decay into right-handed neutrinos is then discussed. Section
\ref{sec:nudecay} computes the lifetime of the right-handed
neutrinos assuming minimal flavor violation and discusses its
dependence on neutrino mass parameters and mixing angles. I conclude
in Section \ref{sec:discussion} with some comments on model-building
issues and summarize results.

\section{Higgs Boson Decay}
\label{sec:higgsdecay}

The renormalizable Lagrangian describing interactions between the
Higgs doublet $H$ $({\bf 1,2})_{-1/2}$, the lepton $SU(2)_W$
doublets $L_i$  $({\bf 1,2})_{-1/2}$,
and three right-handed neutrinos $N_I$ $({\bf 1,1})_{0}$ is given by
\begin{eqnarray}
&& {\cal L_{R}} = \frac{1}{2} m_R N N + \lambda_{\nu} \tilde{H} N L+
\lambda_l H L e^c \label{nulagrangian}
\end{eqnarray} where flavor
indices have been suppressed and $\tilde{H} \equiv i \tau_2 H^*$
where $H$ has a vacuum expectation value (vev) $\langle H \rangle =
v /\sqrt{2}$ and $v \simeq 247$ GeV. Two-component notation is used
throughout this note. We can choose a basis where the Majorana mass
matrix  $m_R$ is diagonal and real with elements $M_I$. In general
they will be non-universal. It will also be convenient to define the
$3 \times 3$ Dirac neutrino mass $m_D \equiv \lambda_{\nu}
v/\sqrt{2}$. The standard see-saw mechanism introduces mass mixing
between the right-handed and left-handed neutrinos which leads to
the active neutrino mass matrix
\begin{eqnarray}
&& m_L = \frac{1}{2}\lambda^T_{\nu} m^{-1}_R \lambda _{\nu} v^2 =
m_D ^T m^{-1}_R m_D ~.
\end{eqnarray}
This is diagonalized by the PMNS matrix $U_{PMNS}$ \cite{PMNS} to
obtain the physical masses $m_I$ of the active neutrinos. At leading
order in the Dirac masses the mass mixing between the left-handed
neutrinos $\nu_I$ and right-handed neutrinos $N_J$ is given by \beq
V_{IJ} =[m^T_D m^{-1}_R ]_{IJ}=[m^T_D]_{IJ} M^{-1}_J \eeq and are
important for the phenomenology of the right-handed neutrinos. For
generic Dirac and Majorana neutrino masses no simple relation exists
between the physical masses, left-right mixing angles and the PMNS
matrix. An estimate  for the neutrino couplings is \beq
 f_I \simeq 7 \times 10^{-7} \left(\frac{m_I}{0.5
\hbox{eV}}\right)^{1/2} \left(\frac{M}{30 \hbox{GeV}}\right)^{1/2}~.
\eeq where $\lambda_{\nu} = U_R f U_L$ has been expressed in terms
of two unitary matrices $U_{L/R}$ and a diagonal matrix $f$ with
elements $f_I$. In general $U_L \neq U_{PMNS}$. Similarly, an
approximate relation for the left-right mixing angles is
\begin{eqnarray} V_{IJ} \simeq
\sqrt{\frac{m_J}{M}} [U_{PMNS}]_{JI}=4 \times
10^{-6}\sqrt{\left(\frac{m_J}{0.5 \hbox{eV}}\right)\left( \frac{30
\hbox{GeV}}{M}\right)} [U_{PMNS}]_{JI}~
\end{eqnarray}
which is valid for approximately universal right-handed neutrino
masses $M_I \simeq M$ and $U_R \simeq 1$. I note that these formulae
for the masses and mixing angles are exact in the limit of universal
Majorana masses and no $CP$ violation in the Dirac masses
\cite{wise:leptonviolation}. For these fiducial values of the
parameters no limits exist from the neutrinoless double $\beta$
decay experiments or collider searches \cite{RHNEW3} because the
mixing angles are too tiny. No limits from cosmology exist either
since the right-handed neutrinos decay before big bang
nucleosynthesis if $M_I \gtap O($GeV$)$, which will be assumed
throughout (see Section \ref{sec:nudecay} for the decay length of
the right-handed neutrinos).

If a right-handed neutrino is lighter than the Higgs boson, $M_I <
m_h$, where $m_h$ is the mass of the Higgs boson, then in principle
there may be new decay channels
\begin{eqnarray}
h \rightarrow N_I +X
\end{eqnarray}
where $X$ may be a Standard Model particle or another right-handed
neutrino (in the latter case $M_I +M_J < m_h$). For instance, from
the neutrino coupling one has $ h \rightarrow N_I \nu_L$. This decay
is irrelevant, however, for practical purposes since the rate is too
small.

But if it is assumed that at the TeV scale there are new dynamics
responsible for generating both the Higgs boson mass and the
right-handed neutrino masses, then higher-dimension operators
involving the two particles should exist and be suppressed by the
TeV scale. These can be a source of new and {\it relevant} decay
processes. Consider then
\begin{eqnarray}
\delta {\cal L}_{eff} &=& \sum_i \frac{c^{(5)}_i \cdot}{\Lambda}
{\cal O}^{(5)}_i  + \sum_i \frac{c^{(6)}_i \cdot}{\Lambda^2} {\cal
O}^{(6)}_i + \cdots + \hbox{h.c.}
\end{eqnarray}
where $\Lambda \simeq {\cal O}($TeV$)$. Only dimension 5 operators
are considered here, with dimension 6 operators discussed elsewhere
\cite{mg2}. The central dot `$\cdot$' denotes a contraction of
flavor indices.

At dimension 5 there are several operators involving right-handed
neutrinos. However it is shown below that constraints from the
observed scale of the left-handed neutrino masses implies that only
one of them can be relevant. It is
\begin{eqnarray}
{\cal O}^{(5)}_1&=& H^{\dagger} H N N \label{o51}
\end{eqnarray}
where the flavor dependence is suppressed. {\em The important point
is that this operator is not necessarily suppressed by any small
Yukawa couplings.} After electroweak symmetry breaking the only
effect of this operator at tree-level is to shift the masses of the
right-handed neutrinos. Constraints on this operator are therefore
weak (see below).

This operator, however, can have a significant effect on the Higgs
boson. For if
\begin{eqnarray}
&&M_I +M_J < m_h~,
\end{eqnarray}
the decay
\begin{eqnarray}
&&h \rightarrow N_I N_J
\end{eqnarray}
can occur. For instance, if only a single flavor is lighter than the
Higgs boson, the decay rate is
\begin{eqnarray}
\Gamma(h \rightarrow N_I N_I) &=&\frac{v^2}{4 \pi \Lambda^2} m_h
\beta_I \left( (\hbox{Re} c^{(5)}_1)^2 \beta^2_I +(\hbox{Im}
c^{(5)}_1)^2 \right)
\end{eqnarray}
where only half the phase space has been integrated over,
$c^{(5)}_1/\Lambda$ is the coefficient of (\ref{o51}), and $\beta_I
\equiv (1- 4 M^2_I /m^2_h)^{1/2}$ is the velocity of the
right-handed neutrino.

The dependence of the decay rate on $\beta$ may be understood from
the following comments. The uninterested reader may skip this
paragraph, since this particular dependence is only briefly referred
to later in the next paragraph, and is not particularly crucial to
any other discussion. Imagine a scattering experiment producing the
two Majorana fermions only through an on-shell Higgs boson in the
$s$-channel. The cross-section for this process is related to the
decay rate into this channel, and in particular their dependence on
the final state phase space are identical. Conservation of angular
momentum, and when appropriate, conservation of $CP$ in the
scattering process then fixes the dependence of $\Gamma$ on phase
space. For example, note that the phase of $c^{(5)}_1$ is physical
and cannot be rotated away. When $\hbox{Im} c^{(5)}_1=0$ the
operator (\ref{o51}) conserves $CP$ and the decay rate has the
$\beta^3$ dependence typical for fermions. This dependence follows
from the usual argument applied to Majorana fermions: a pair of
Majorana fermions has an intrinsic $CP$ parity of $-1$
\cite{weinbergbook}, so conservation of $CP$ and total angular
momentum in the scattering process implies that the partial wave
amplitude for the two fermions must be a relative $p$-wave state. If
the phase of $c^{(5)}_1$ is non-vanishing, then $CP$ is broken and
the partial wave amplitude can have both $p$-wave and $s$-wave
states while still conserving total angular momentum. The latter
amplitude leads to only a $\beta_I$ phase space suppression.

There is a large region of parameter space where this decay rate is
larger than the rate for the Higgs boson to decay into bottom
quarks, and, if kinematically allowed, not significantly smaller
than the rate for the Higgs boson to decay into electroweak gauge
bosons. For example, with $\hbox{Im}(c^{(5)}_1)=0$ and no sum over
$I$, \beq \frac{\Gamma(h \rightarrow N_I N_I)}{\Gamma(h \rightarrow
b \overline{b})} =\frac{2(c^{(5)}_1)^2}{3} \frac{v^4}{m^2_b
\Lambda^2} \beta^3_I \eeq This ratio is larger than 1 for $\Lambda
 \ltap 12 |c^{(5)}_1| \beta^{3/2}_I $ TeV .
If all three right-handed neutrinos are lighter than the Higgs
boson, then the total rate into these channels is larger than the
rate into bottom quarks for $\Lambda \ltap 20 |c^{(5)}_1|
\beta^{3/2}_I$ TeV. If $\hbox{Im}(c^{(5)}_1) \neq 0$ the operator
violates $CP$ and the region of parameter space where decays to
right-handed neutrinos dominate over decays to bottom quarks becomes
larger, simply because now the decay rate has less of a phase space
suppression, as described above. The reason for the sensitivity to
large values of $\Lambda$ is because the bottom Yukawa coupling is
small. For
\begin{eqnarray}
&&m_h > 2 m_W
\end{eqnarray}
the Higgs boson can decay into a pair of $W$ bosons with a large
rate and if kinematically allowed, into a pair of $Z$ gauge bosons
with a branching ratio of approximately $1/3$. One finds that with
$\hbox{Im}(c^{(5)}_1)=0$ and no sum over $I$, \beq \frac{\Gamma(h
\rightarrow N_I N_I)}{\Gamma(h \rightarrow WW)} =\frac{4
(c^{(5)}_1)^2 v^4}{m^2_h \Lambda^2}
\frac{\beta^3_I}{\beta_W}\frac{1}{f(\beta_W)} \eeq where
$f(\beta_W)=3/4 -\beta^2_W/2+ 3 \beta^4_W/4$ \cite{Higgsguide} and
$\beta_W$ is the velocity of the $W$ boson. Still, the decay of the
Higgs boson into right-handed neutrinos is not insignificant. For
example, with $\Lambda \simeq 2$ TeV, $c^{(5)}_1=1$ and $\beta_I
\simeq 1$, the branching ratio for a Higgs boson of mass $300$ GeV
to decay into a single right-handed neutrino flavor of mass $30$ GeV
is approximately $5\%$. Whether the decays of the Higgs boson into
right-handed neutrinos are visible or not depends on the lifetime of
the right-handed neutrino. That issue is discussed in Section
\ref{sec:nudecay}.

It is now shown that all the other operators at $d=5$ involving
right-handed neutrinos and Higgs bosons are irrelevant for the decay
of the Higgs boson. Aside from (\ref{o51}), there is only one more
linearly independent operator involving the Higgs boson and a
neutrino,
\begin{eqnarray}
&&{\cal O}^{(5)}_2 = L \tilde{H} L \tilde{H}~. \label{d5numass}
\end{eqnarray}
After electroweak symmetry breaking this operator contributes to the
left-handed neutrino masses, so its coefficient must be tiny,
$c^{(5)}_2 v^2 /\Lambda \ltap O(m_{\nu_L}) ~.$ Consequently, the
decay of the Higgs boson into active neutrinos from this operator is
irrelevant. In Section \ref{sec:minimalflavorviolation} it is seen
that under the minimal flavor violation hypothesis this operator is
naturally suppressed to easily satisfy the condition above. It is
then consistent to assume that the dominant contribution to the
active neutrino masses comes from mass mixing with the right-handed
neutrinos.

Other operators involving the Higgs boson exist at dimension 5, but
all of them can be reduced to (\ref{d5numass}) and dimension 4
operators by using the equations of motion. For instance,
\begin{eqnarray}
 {\cal O}^{(5)}_3 &\equiv& - i(\partial ^{\mu}
\overline{N}) \overline{\sigma}^{\mu} L \tilde{H}  \rightarrow m_R N
L \tilde{H} + (\tilde{H} L) \lambda^T_{\nu}  (L \tilde{H}) ~,
\label{eomo53}
\end{eqnarray}
where the equations of motion were used in the last step. As a
result, this operator does not introduce any new dynamics. Still,
its coefficients must be tiny enough to not generate too large of a
neutrino mass. In particular, enough suppression occurs if its
coefficients are less than or comparable to the neutrino couplings.
Under the minimal flavor violation hypothesis it is seen that these
coefficients are naturally suppressed to this level.

Even if the operators ${\cal O}^{(5)}_2$ and ${\cal O}^{(5)}_3$ are
not present at tree-level, they will be generated at the loop-level
through operator mixing with ${\cal O}^{(5)}_1$. This is because the
overall lepton number symmetry $U(1)_{LN}$ is broken with both the
neutrino couplings and ${\cal O}^{(5)}_1$ present. However, such
mixing will always involve the neutrino couplings and be small
enough to not generate too large of a neutrino mass. To understand
this, it is useful to introduce a different lepton number under
which the right-handed neutrinos are neutral and both the charged
leptons and left-handed neutrinos are charged. Thus the neutrino
couplings and the operators ${\cal O}^{(5)}_2$ and ${\cal
O}^{(5)}_3$ violate this symmetry, but the operator ${\cal
O}^{(5)}_1$ preserves it. In the limit that $\lambda_{\nu}
\rightarrow 0$ this lepton number symmetry is perturbatively exact,
so inserting ${\cal O}^{(5)}_1$ into loops can only generate ${\cal
O}^{(5)}_2$ and ${\cal O}^{(5)}_3$ with coefficients proportional to
the neutrino couplings. Further, ${\cal O}^{(5)}_2$ violates this
symmetry by two units, so in generating it from loops of Standard
Model particles and insertions of ${\cal O}^{(5)}_1$ it will be
proportional to at least two powers of the neutrino couplings.
Likewise, in generating ${\cal O}^{(5)}_3$ from such loops its
coefficient is always proportional to at least one power of the
neutrino coupling. In particular, ${\cal O}^{(5)}_2$ is generated
directly at two-loops, with $c^{(5)}_2 \propto \lambda^T _{\nu}
\lambda_{\nu} c^{(5)}_1$. It is also generated indirectly at
one-loop, since ${\cal O}^{(5)}_3$ is generated at one-loop, with
$c^{(5)}_3 \propto c^{(5)}_1 \lambda_{\nu}$. These operator mixings
lead to corrections to the neutrino masses that are suppressed by
loop factors and at least one power of $m_R / \Lambda$ compared to
the tree-level result.

As a result, no significant constraint can be applied to the
operator ${\cal O}^{(5)}_1$.\footnote{This statement assumes
$c^{(5)} \ltap O(16 \pi^2)$ and that the loop momentum cutoff
$\Lambda_{\rm loop} \simeq \Lambda$. Constraints might conceivably
occur for very light right-handed neutrino masses, but that
possibility is not explored here since $M_I \gtap O($GeV$)$ is
assumed throughout in order that the right-handed neutrinos decay
before big bang nucleosynthesis.} Instead the challenge is to
explain why the coefficients of ${\cal O}^{(5)}_2$ and ${\cal
O}^{(5)}_3$ in the effective theory are small to begin with. The
preceding arguments show why it is technically natural for them to
be small, even if ${\cal O}^{(5)}_1$ is present. The minimal flavor
violation hypothesis discussed below does provide a technically
consistent framework in which this occurs.

\section{Naturalness}
\label{sec:naturalness}

The operator \beq \frac{c^{(5)}_1}{\Lambda} H^{\dagger} H N N \eeq
violates chirality, so it contributes to the mass of the
right-handed neutrino at both tree and loop level. At tree level
\beq \delta m_R = c^{(5)}_1 \frac{v^2}{\Lambda} = 60 c^{(5)}_1
\left(\frac{\hbox{TeV}}{\Lambda} \right) \hbox{GeV} ~.\eeq There is
also a one-loop diagram with an insertion of this operator. It has a
quadratic divergence such that  \beq \delta m_R \simeq 2 c^{(5)}_1
\frac{\Lambda}{16 \pi^2}~. \eeq Similarly, at one-loop \beq \delta
m^2_h \simeq  \frac{1}{16 \pi^2} \hbox{Tr}[c^{(5)}_1 m_R] \Lambda ~.
\eeq If $c^{(5)}_1 \sim O(1)$ then a right-handed neutrino with mass
$M_I \simeq 30$ GeV requires $ O(1)$ tuning for TeV $\lsim \Lambda
\lsim 10$ TeV, and $m_h \simeq 100$ GeV is technically natural
unless $\Lambda \gsim 10$ TeV or $m_R$ is much larger than the range
$(M_I \ltap 150$ GeV $)$ considered here.

Clearly, if $\Lambda \gsim O(10$ TeV$)$ then a symmetry would be
required to protect the right-handed neutrino and Higgs boson
masses. One such example is supersymmetry. Then this operator can be
generalized to involve both Higgs superfields and would appear in
the superpotential. It would then be technically natural for the
Higgs boson and right-handed neutrino masses to be protected, even
for large values of $\Lambda$. As discussed previously, for such
large values of $\Lambda$ decays of the Higgs boson into
right-handed neutrinos may still be of phenomenological interest.

\section{Minimal Flavor Violation}
\label{sec:minimalflavorviolation}

The higher dimension operators involving right-handed neutrinos and
Standard Model leptons previously discussed can {\em a priori} have
an arbitrary flavor structure and size. But as is well-known, higher
dimension operators in the lepton and quark sector suppressed by
only $\Lambda \simeq $ TeV $-10$ TeV are grossly excluded by a host
of searches for flavor changing neutral currents and overall lepton
number violating decays.

A predictive framework for the flavor structure of these operators
is provided by the minimal flavor violation hypothesis
\cite{mfvprinciple,mfvprinciple2,wise:leptonviolation}. This
hypothesis postulates a flavor symmetry assumed to be broken by a
minimal set of non-dynamical fields, whose vevs determine the
renormalizable Yukawa couplings and masses that violate the flavor
symmetry. Since a minimal field content is assumed, the flavor
violation in higher dimension operators is completely determined by
the now {\em irreducible} flavor violation appearing in the
right-handed neutrino masses and the neutrino, charged lepton and
quark Yukawa couplings. Without the assumption of a minimal field
content breaking the flavor symmetries, unacceptably large flavor
violating four fermion operators occur. In practice, the flavor
properties of a higher dimension operator is determined by inserting
and contracting appropriate powers and combinations of Yukawa
couplings to make the operator formally invariant under the flavor
group. Limits on operators in the quark sector are $5-10$ TeV
\cite{mfvprinciple2}, but weak in the lepton sector unless the
neutrinos couplings are not much less than order unity
\cite{wise:leptonviolation}\cite{grinstein}.

It is important to determine what this principle implies for the
size and flavor structure of the operator \beq
(c^{(5)}_1)_{IJ}H^{\dagger} H N_I N_J ~. \label{o51mv} \eeq It is
seen below that the size of its coefficients depends critically on
the choice of the flavor group for the right-handed neutrinos. This
has important physical consequences which are then discussed.

In addition one would like to determine whether the operators ${\cal
O}^{(5)}_2$ and ${\cal O}^{(5)}_3$ are sufficiently suppressed such
that their contribution to the neutrinos masses is always
subdominant. In Section \ref{sec:higgsdecay} it was argued that if
these operators are initially absent, radiative corrections
involving ${\cal O}^{(5)}_1$ and the neutrino couplings will never
generate large coefficients (in the sense used above) for these
operators. However, a separate argument is needed to explain why
they are initially small to begin with. It is seen below that this
is always the case assuming minimal flavor violation.

To determine the flavor structure of the higher dimension operators
using the minimal flavor violation hypothesis, the transformation
properties of the particles and couplings are first defined. The
flavor symmetry in the lepton sector is taken to be
\begin{eqnarray} && G_N \times SU(3)_{L} \times
SU(3)_{e^c}\times U(1)
\end{eqnarray}
where $U(1)$ is the usual overall lepton number acting on the
Standard Model leptons. With right-handed neutrinos present there is
an ambiguity over what flavor group to choose for the right-handed
neutrinos, and what charge to assign them under the $U(1)$. In fact,
since there is always an overall lepton number symmetry unless both
the Majorana masses and the neutrino couplings are non-vanishing,
there is a maximum of two such $U(1)$ symmetries.

Two possibilities are considered for the flavor group of the
right-handed neutrinos: \beq G_N = SU(3) \times U(1)^{\prime}
~\hbox{or}~ SO(3) ~. \eeq The former choice corresponds to the
maximal flavor group, whereas the latter is chosen to allow for a
large coupling for the operator (\ref{o51mv}), shown below. The
fields transform under the flavor group $SU(3) \times SU(3)_L \times
SU(3)_{e^c} \times U(1)^{\prime} \times U(1)$ as
\begin{eqnarray}
N & \rightarrow & ({\bf 3},{\bf 1}, {\bf 1})_{({\bf 1},{\bf 0})} \\
L & \rightarrow &  ({\bf 1},{\bf 3}, {\bf 1})_{({\bf -1},{\bf 1})}\\
e^c & \rightarrow & ({\bf 1},{\bf 1}, {\bf 3})_{({\bf 1},{\bf -1})}
~,
\end{eqnarray}
Thus $U(1)^{\prime}$ is a lepton number acting on the right-handed
neutrinos and Standard Model leptons and is broken only by the
Majorana masses. $U(1)$ is a lepton number acting only on the
Standard Model leptons and is only broken by the neutrino couplings.
Then the masses and Yukawa couplings of the theory are promoted to
spurions transforming under the flavor symmetry. Their
representations are chosen in order that the Lagrangian is formally
invariant under the flavor group. Again for $G_N=SU(3)\times
U(1)^{\prime}$,
\begin{eqnarray}
\lambda_{\nu} & \rightarrow & ({\bf \overline{3}},
\bf{\overline{3}},\bf{1})_{({\bf 0},{\bf -1})}   \\
\lambda_l & \rightarrow & (\bf{1},\bf{\overline{3}},
\bf{\overline{3}})_{({\bf 0},{\bf 0})} \\
m_R & \rightarrow & ({\bf \overline{6}}, \bf{1},\bf{1})_{({\bf
-2},{\bf 0})}~.
\end{eqnarray}
For $G_N=SO(3)$ there are several differences. First, the ${\bf
\overline{3}}$'s of $SU(3)$ simply become ${\bf 3}$'s of $SO(3)$.
Next, the $U(1)$ charge assignments remain but there is no
$U(1)^{\prime}$ symmetry. Finally, a minimal field content is
assumed throughout, implying that for $G_N=SO(3)$ $m_R \sim {\bf 6}$
is real.

With these charge assignments a spurion analysis can now be done to
estimate the size of the coefficents of the dimension 5 operators
introduced in Section \ref{sec:higgsdecay}.

For either choice of $G_N$ one finds the following. An operator that
violates the $U(1)$ lepton number by $n$ units is suppressed by $n$
factors of the tiny neutrino couplings. In particular, the dangerous
dimension 5 operators ${\cal O}^{(5)}_2$ and ${\cal O}^{(5)}_3$ are
seen to appear with two and one neutrino couplings, which is enough
to suppress their contributions to the neutrino masses. If
$G_N=SO(3)$ such operators can also be made invariant under $SO(3)$
by appropriate contractions. If however $G_N = SU(3) \times
U(1)^{\prime}$, then additional suppressions occur in order to
construct $G_N$ invariants. For example, the coefficients of the
dimension 5 operators ${\cal O}^{(5)}_2$ and ${\cal O}^{(5)}_3$ are
at leading order $\lambda^T_{\nu} m^{\dagger}_R
\lambda_{\nu}/\Lambda$ and $\lambda_{\nu} m^{\dagger}_R/\Lambda$
respectively and are sufficiently small.

It is now seen that the flavor structure of the operator (\ref{o51})
depends on the choice of the flavor group $G_N$. One finds
\begin{eqnarray}
G_N = SU(3)\times U(1)^{\prime} &:& c^{(5)}_1 \sim a_1
\frac{{m_R}}{\Lambda} + a_2 \frac{{m_R  \hbox{Tr}[m^{\dagger}_R
m_R]}}{\Lambda^2} +\cdots
 \nonumber \\
G_N= SO(3) &:& c^{(5)}_1 \sim {\bf 1} + d_1 \frac{m_R}{\Lambda} +
d_2 \frac{{m_R \cdot m_R}}{\Lambda^2} + \cdots + e_1 \lambda_{\nu}
\lambda^{\dagger}_{\nu} + \cdots \label{gnexpression}
\end{eqnarray} where $\cdots$ denotes higher powers in $m_R$ and
$\lambda_{\nu} \lambda^{\dagger}_{\nu}$. Comparing the expressions
in (\ref{gnexpression}), the only important difference between the
two is that ${\bf 1}$ is invariant under $SO(3)$, but not under
$SU(3)$ or $U(1)^{\prime}$. As we shall see shortly, this is a key
difference that has important consequences for the decay rate of the
Higgs boson into right-handed neutrinos.

Next the physical consequences of the choice of flavor group are
determined. First note that if we neglect the $\lambda _{\nu}
\lambda^{\dagger} _{\nu} \propto m_{L}$ contribution to $c^{(5)}_1$,
then for either choice of flavor group the right-handed neutrino
masses $m_R$ and couplings $c^{(5)}_1$ are simultaneously
diagonalizable. For $G_N=SO(3)$ this follows from the assumption
that $m_R \sim {\bf 6}$ is a real representation. As a result, the
couplings $c^{(5)}_1$ are flavor-diagonal in the right-handed
neutrino mass basis.

If $G_N=SO(3)$ the couplings $c^{(5)}_1$ are {\em flavor-diagonal,
universal at leading order, and not suppressed by any Yukawa
couplings}. It follows that \beq \frac{\hbox{Br}(h \rightarrow N_I
N_I)}{\hbox{Br}(h \rightarrow N_J N_J)} =
\frac{\beta^3_I}{\beta^3_J} \simeq 1\eeq up to small flavor-diagonal
corrections of order $m_R/\Lambda$ from the next-to-leading-order
terms in the couplings $c^{(5)}_1$. $\beta_I$ is the velocity of
$N_I$ and its appearance in the above ratio is simply from phase
space. It is worth stressing that even if the right-handed neutrino
masses are non-universal, the branching ratios of the Higgs boson
into the right-handed neutrinos are approximately universal and
equal to $1/3$ up to phase space corrections. The calculations from
Section \ref{sec:higgsdecay} of the Higgs boson decay rate into
right-handed neutrinos do not need to be rescaled by any small
coupling, and the conclusion that these decay channels dominate over
$h \rightarrow b \overline{b}$ for $\Lambda$ up to 20 TeV still
holds. Theoretically though, the challenge is to understand why $M_I
\ll \Lambda$.

Similarly, if $G_N=SU(3)$ the couplings are {\em flavor-diagonal and
suppressed by at least a factor of $m_R/\Lambda$ but not by any
Yukawa couplings.} This suppression has two effects. First, it
eliminates the naturalness constraints discussed in Section
\ref{sec:naturalness}. The other is that it suppresses the decay
rate of $h \rightarrow N_I N_I$ by a predictable amount. In
particular \begin{eqnarray} \Gamma(h \rightarrow N_I N_I)
&=&\frac{v^2}{4 \pi \Lambda^2}\left(\frac{M_I}{\Lambda}\right)^2 m_h
\beta^3_I
\end{eqnarray}
where I have set $a_1=1$, and
 \beq \frac{\hbox{Br}(h \rightarrow N_I N_I)} {\hbox{Br}(h
\rightarrow N_J N_J)} = \frac{M^2_{I}}{M^2_{J}}
\frac{\beta^3_I}{\beta^3_J} \eeq up to flavor-diagonal corrections
of order $m_R/\Lambda$. In this case, the Higgs boson decays
preferentially to the right-handed neutrino that is the heaviest.
Still, even with this suppression these decays dominate over $h
\rightarrow b \overline{b}$ up to $\Lambda \simeq 1$ TeV if three
flavors of right-handed neutrinos of mass $M_I \simeq O(50$GeV$)$
are lighter than the Higgs boson. For larger values of $\Lambda$
these decays have a subdominant branching fraction. They are still
interesting though, because they have a rich collider phenomenology
and may still be an important channel in which to search for the
Higgs boson. This scenario might be more natural theoretically,
since an approximate $SU(3)$ symmetry is protecting the mass of the
fermions.

\section{Right-handed Neutrino Decays} \label{sec:nudecay}

I have discussed how the presence of a new operator at the TeV scale
can introduce new decay modes of the Higgs boson into lighter
right-handed neutrinos, and described the circumstances under which
these new processes may be the dominant decay mode of the Higgs
boson. In the previous section we have seen that whether that
actually occurs or not depends critically on a few assumptions. In
particular, on whether the Higgs boson is light, on the scale of the
new operator, and key assumptions about the identity of the broken
flavor symmetry of the right-handed neutrinos.

Whether the decays of the Higgs boson into right-handed neutrinos
are visible or not depends on the lifetime of the right-handed
neutrinos. It is seen below that in the minimal flavor violation
hypothesis their decays modes are determined by their renormalizable
couplings to the electroweak neutrinos and leptons, rather than
through higher-dimension operators.

The dominant decay of a right-handed neutrinos is due to the gauge
interactions with the electroweak gauge bosons it acquires through
mass mixing with the left-handed neutrinos. At leading order a
right-handed neutrino $N_J$ acquires couplings to $W l_I$ and $Z
\nu_I$ which are identical to those of a left-handed neutrino,
except that they are suppressed by the mixing angles \beq V_{IJ} =
[m^T _D]_{IJ} M^{-1}_{J} ~. \eeq

If the right-handed neutrino is heavier than the electroweak gauge
bosons but lighter than the Higgs boson, it can decay as $N_J
\rightarrow W^+ l^-_I$ and $N_J \rightarrow Z \nu_I$. Since it is a
Majorana particle, decays to charge conjugated final states also
occur. The rate for these decays is proportional to $|V_{IJ}|^2
M^3_J$.

If a right-handed neutrino is lighter than the electroweak gauge
bosons, it decays through an off-shell gauge boson to a three-body
final state. Its lifetime can be obtained by comparing it to the
leptonic decay of the $\tau$ lepton, but after correcting for some
additional differences described below. The total decay rate is
\footnote{An $\approx 2$ error in an earlier version has been
corrected.} \beq \frac{\Gamma_{\hbox{total}}(N_I)}{\Gamma(\tau
\rightarrow \mu \overline{\nu}_{\mu} \nu_{\tau})} = 2 \times 9
\left(c_W +0.40 c_Z\right) \frac{[m_D  m^{\dagger}_D ]_{II}}{M^2_I}
\left(\frac{M_I}{m_{\tau}}\right)^5 ~. \label{decaylength} \eeq The
corrections are the following. The factor of $``9"$ counts the
number of decays available to the right-handed neutrino through
charged current exchange, assuming it to be heavier than roughly
few-10 GeV. The factor of $``0.40"$ counts the neutral current
contribution. It represents about $30\%$ of the branching ratio,
with the remaining $70\%$ of the decays through the charged current.
The factor of $``2"$ is because the right-handed neutrino is a
Majorana particle, so it can decay to both particle and
anti-particle, e.g. $W^* l^-$ and $W^* l^+$, or $Z^* \nu$ and $Z^*
\overline{\nu}$. Another correction is due to the finite momentum
transfer in the electroweak gauge boson propagators. This effect is
described by the factors $c_W$ and $c_W$ where \beq c_G(x_G,y_G) = 2
\int^1 _0 dz z^2(3-2z) \left((1-(1-z)x_G)^2+ y_G \right)^{-1}
\label{cG} \eeq where $x_G = M^2_I/m^2_G$, $y_G = \Gamma^2_G/m^2_G$,
$c_G(0,0)=1$ and each propagator has been approximated by the
relativistic Breit-Wigner form. The non-vanishing momentum transfer
enhances the decay rate by approximately $10\%$ for $m_R$ masses
around $30 \hbox{GeV}$ and by approximately $50\%$ for masses around
$50$ GeV. This effect primarily affects the overall rate and is less
important to the individual branching ratios.

The formula (\ref{cG}) is also valid when the right-handed neutrino
is more massive than the electroweak gauge bosons such that the
previously mentioned on-shell decays occur. In that case
(\ref{decaylength}) gives the inclusive decay rate of a right-handed
neutrino into any electroweak gauge boson and a charged lepton or a
left-handed neutrino. In this case the correction from the momentum
transfer is obviously important to include! It enhances the decay
rate by approximately a factor of $40$ for masses around $100$ GeV,
but eventually scales as $M^{-2}_I$ for a large enough mass.

An effect not included in the decay rate formula above is the
quantum interference that occurs in the same flavor $l^+ l^- \nu$ or
$\nu \nu \overline{\nu}$ final states. Its largest significance is
in affecting the branching ratio of these specific, subdominant
decay channels and is presented elsewhere \cite{mg2}. Using $c
\tau_{\tau} = 87 \mu m $ \cite{PDG} and $BR(\rightarrow \mu
\overline{\nu}_{\mu} \nu_{\tau})=0.174$ \cite{PDG},
(\ref{decaylength}) gives the following decay length for $N_I$, \beq
c \tau _I = 0.90 m \left(\frac{1.40}{c_W + 0.40 c_Z}\right)
\left(\frac{\hbox{30 GeV}}{M_I}
 \right)^3 \left(\frac{(120 \hbox{ keV})^2}{[ m_D m^{\dagger}_D]_{II}}
 \right)~.
\eeq Care must be used in interpreting this formula, since the Dirac
and Majorana masses are not completely independent because they must
combine together to give the observed values of the active neutrino
masses.

This expression is both model-independent and model-dependent. Up to
this point no assumptions have been made about the elements of the
Dirac mass matrix or the right-handed neutrino masses, so the result
above is completely general. Yet the actual value of the decay
length clearly depends on the flavor structure of the Dirac mass
matrix. In particular, the matrix elements $[m_D
m^{\dagger}_D]_{II}/M_I$ are not the same as the active neutrino
mass masses. This is fortunate, since it presents an opportunity to
measure a different set of neutrino parameters from those measured
in neutrino oscillations.

The masses $M_I$ describe 3 real parameters, and {\em a priori} the
Dirac matrix $m_D$ describes 18 real parameters. However, 3 of the
phases in $m_D$ can be removed by individual lepton number phase
rotations on the left-handed neutrinos and charged leptons, leaving
15 parameters which I can think of as 6 mixing angles, 3 real Yukawa
couplings and 6 phases. Including the three right-handed neutrino
masses gives 18 parameters in total. Five constraints on
combinations of these 18 parameters already exist from neutrino
oscillation experiments. In principle all of these parameters could
be measured through detailed studies of right-handed neutrino
decays, since amplitudes for individual decays are proportional to
the Dirac neutrino matrix elements. However, at tree-level these
observables depend only on $|[m_{D}]_{IJ}|$ and are therefore
insensitive to the 6 phases. So by studying tree-level processes
only the 3 right-handed neutrino masses, 3 Yukawa couplings, and 6
mixing angles could be measured in principle.

In particular, the dominant decay is $h \rightarrow N_I N_I
\rightarrow q q q q  l_J l_K$ which contains no missing energy.
Since the secondary events are highly displaced, there should be no
confusion about which jets to combine with which charged leptons. In
principle a measurement of the mass of the right-handed neutrino and
the Higgs boson is possible by combining the invariant momentum in
each event. A subsequent measurement of a right-handed neutrino's
lifetime from the spatial distribution of its decays measures $[m_D
m^{\dagger}_D]_{II}$. More information is acquired by measuring the
nine branching ratios $BR(N_I \rightarrow q q^{\prime} l_J) \propto
|[m_D]_{IJ}|^2$. Such measurements provide 6 additional independent
constraints. In total, 12 independent constraints on the 18
parameters could in principle be obtained from studying right-handed
neutrino decays at tree-level.

To say anything more precise about the decay length would require a
model of the neutrino couplings and right-handed neutrino mass
parameters. Specific predictions could be done within the context of
such a model. Of interest would be the branching ratios and the mean
and relative decay lengths of the three right-handed neutrinos.

The factor $[m_D m^{\dagger}_D]_{II}/M_I$ appearing in the decay
length is not the active neutrino mass obtained by diagonalizing
$m^T_D m^{-1}_R m_D$, but it is close. If I approximate $[m_D
m^{\dagger}_D]_{II}/M_I \simeq m_{I}$, then \beq c \tau_I \simeq
0.90 m \left(\frac{30 \hbox{ GeV}}{M_I} \right)^4\left(\frac{0.48
\hbox{ eV}}{m_I} \right) \left(\frac{1.40}{c_W+0.4 c_Z} \right)\eeq
A few comments are in order. First, the decay lengths are
macroscopic, since by inspection they range from $O(100\mu m)$ to
$O(10m)$ for a range of parameters, and for these values are
therefore visible at colliders. Next, the decay length is evidently
extremely sensitive to $M_I$. Larger values of $M_I$ have shorter
decays lengths. For instance, if $M_I=100$ GeV (which requires $m_h
>200$ GeV) and $m_I=0.5$ eV then $c \tau_I \simeq 0.2 mm$.
Finally, if the active neutrino masses are hierarchical, then one
would expect $ M^4_I c \tau_I$ to be hierarchical as well, since
this quantity is approximately proportional to $m^{-1}_{L}$. One or
two right-handed neutrinos may therefore escape the detector if the
masses of the lightest two active neutrinos are small enough.

I have described decays of the right-handed neutrinos caused by its
couplings to electroweak gauge bosons acquired through mass mixing
with the left-handed neutrinos. However, additional decay channels
occur through exchange of an off-shell Higgs boson, higher dimension
operators or loop effects generated from its gauge couplings. It
turns out that these processes are subdominant, but may be of
interest in searching for the Higgs boson. Exchange of an off-shell
Higgs boson causes a decay $N_I \rightarrow \nu_J b \overline{b}$
which is suppressed compared to the charged and neutral current
decays by the tiny bottom Yukawa coupling. Similarly, the dimension
5 operator (\ref{o51}) with generic flavor couplings allows for the
decay $N_I \rightarrow N_J h^{*} \rightarrow N_J b \overline{b}$ for
$N_J$ lighter than $N_I$ \footnote{The author thanks Scott Thomas
for this observation.}. However, using the minimal flavor violation
hypothesis it was shown in Section \ref{sec:minimalflavorviolation}
that the couplings of that higher dimension operator are diagonal in
the same basis as the right-handed neutrino mass basis, up to
flavor-violating corrections that are at best $O(\lambda^2_{\nu})$
(see (\ref{gnexpression})). As result, this decay is highly
suppressed. At dimension 5 there is one more operator that I have
not yet introduced which is the magnetic moment operator \beq
\frac{c^{(5)}_4}{\Lambda} \cdot N \sigma ^{\rho \sigma} N B_{\rho
\sigma}
 \eeq
and it involves only two right-handed neutrinos. It causes a heavier
right-handed neutrino to decay into a lighter one,  $N_I \rightarrow
N_J + \gamma/Z$ for $I \neq J$. To estimate the size of this
operator, first note that its coefficient must be anti-symmetric in
flavor. Then in the context of minimal flavor violation with
$G_R=SO(3)$, the leading order term is $ c^{(5)}_4 \simeq
[\lambda_{\nu} \lambda^{\dagger}_{\nu}]_{AS} $ where ``AS" denotes
`anti-symmetric part'. This vanishes unless the neutrino couplings
violate $CP$. In that case the amplitude for this decay is of order
$(\lambda_{\nu})^2$. If $G_R =SU(3) \times U(1)^{\prime}$ the
leading order term cannot be $[m_R]_{AS}(\hbox{Tr}[m_R
m^{\dagger}_R]^q)^n/\Lambda^{n+q}$, since they vanish in the
right-handed neutrino mass basis. The next order involves
$\lambda_{\nu} \lambda^{\dagger}_{\nu}$ and some number of $m_R$'s,
but there does not appear to be any invariant term. Thus for either
choice of $G_R$ the amplitude for $N_I$ decays from this operator
are $O(\lambda^2_{\nu})$ or smaller, which is much tinier than the
amplitudes for the other right-handed neutrino decays already
discussed which are of order $\lambda_{\nu}$.
 Subdominant decays $N \rightarrow \nu + \gamma$ can
occur from dimension 6 operators and also at also one-loop from
electroweak interactions, but in both cases the branching ratio is
tiny \cite{mg2}.

\section{Discussion}
\label{sec:discussion}

In order for these new decays to occur at all requires that the
right-handed neutrinos are lighter than the Higgs boson. But from a
model building perspective, one may wonder why the right-handed
neutrinos are not {\em heavier} than the scale $\Lambda$. A scenario
in which the right-handed neutrinos are composite would naturally
explain why these fermions are comparable or lighter than the
compositeness scale $\Lambda$, assumed to be $O($TeV$)$. Since their
interactions with the Higgs boson through the dimension 5 operator
(\ref{o51}) are not small, the Higgs boson would be composite as
well (but presumed to be light).

These new decay channels of the Higgs boson will be the dominant
decay modes if the right-handed neutrinos are also lighter than the
electroweak gauge bosons, and if the coefficient of the higher
dimension operator (\ref{o51}) is not too small. As discussed in
Section \ref{sec:minimalflavorviolation}, in the minimal flavor
violation framework the predicted size of this operator depends on
the choice of approximate flavor symmetries of the right-handed
neutrinos. It may be $O(1)$ or $O(m_R/\Lambda)$.

In the former situation the new decays dominate over Higgs boson
decays to bottom quarks for scales $\Lambda \ltap 10-20$ TeV,
although only scales $\Lambda \simeq 1 -10$ TeV are technically
natural. This case however presents a challenge to model building,
since the operator (\ref{o51}) breaks the chirality of the
right-handed neutrinos. Although it may be technically natural for
the right-handed neutrinos to be much lighter than the scale
$\Lambda$ (see Section \ref{sec:naturalness}), one might expect that
any theory which generates a large coefficient for this operator to
also generate Majorana masses $m_R \sim O(\Lambda)$.

In the case where the coefficient of (\ref{o51}) is $O(m_R/\Lambda)$
the new decays can still dominate over decays to bottom quarks
provided that the scale $\Lambda \simeq O(1$ TeV$)$. For larger
values of $\Lambda$ these decays are subdominant but have sizable
branching fractions up to $\Lambda \simeq O(10$TeV$)$. This
situation might be more amendable to model building. For here an
approximate $SU(3)$ symmetry is protecting the mass of the
right-handed neutrinos.

In either case though one needs to understand why the right-handed
neutrinos are parametrically lighter than $\Lambda$. It would be
extremely interesting to find non-QCD-type theories of strong
dynamics where fermions with masses parametrically lighter than the
scale of strong dynamics occur. Or using the AdS/CFT correspondence
\cite{adscft}, to find a Randall-Sundrum type model \cite{rs} that
engineers this outcome. The attitude adopted here has been to assume
that such an accident or feature can occur and to explore the
consequences.

Assuming that these theoretical concerns can be naturally
addresseed, the Higgs boson physics is quite rich. To summarize, in
the new process the Higgs boson decays through a cascade into a
six-body or four-body final state depending on the masses of the
right-handed neutrinos. First, it promptly decays into a pair of
right-handed neutrinos, which have a macroscopic decay length
anywhere from $O(100 \mu m-10m)$ depending on the parameters of the
Majorana and Dirac neutrino masses. If one or two active neutrinos
are very light, then the decay lengths could be larger. Decays
occurring in the detector appear as a pair of displaced vertices.
For most of the time each secondary vertex produces a quark pair and
a charged lepton, dramatically violating lepton number. For a
smaller fraction of the time a secondary vertex produces a pair of
charged leptons or a pair of quarks, each accompanied with missing
energy. From studying these decays one learns more about neutrinos
and the Higgs boson, even if these channels should not form the
dominant decay mode of the Higgs boson. The experimental constraints
on this scenario from existing colliders and its discovery potential
at the LHC will be described elsewhere \cite{mg2} \cite{gls}.

\section*{Acknowledgments} The author
thanks Matt Stassler and Scott Thomas for discussions.
This work is supported by the U.S. Department of Energy under
contract No. DE-FG03-92ER40689.

\end{document}